\documentclass{emulateapj}
\usepackage{natbib}

\slugcomment{To appear in ApJ.}
\shorttitle{SN 2012aw Progenitor}
\shortauthors{Van Dyk et al.}

\begin{document}

\title{The Red Supergiant Progenitor of Supernova 2012aw (PTF12bvh) in Messier 95}

\author{Schuyler D.~Van Dyk\altaffilmark{1}, S.~Bradley Cenko\altaffilmark{2},
Dovi Poznanski\altaffilmark{3}, Iair Arcavi\altaffilmark{4},
Avishay Gal-Yam\altaffilmark{4}, Alexei V.~Filippenko\altaffilmark{2},
Kathryn Silverio\altaffilmark{2}, Alan Stockton\altaffilmark{5}, Jean-Charles Cuillandre\altaffilmark{6},
Geoffrey W.~Marcy\altaffilmark{2}, Andrew W. Howard\altaffilmark{2}, and 
Howard Isaacson\altaffilmark{2}}

\altaffiltext{1}{Spitzer Science Center/Caltech, Mailcode 220-6,
  Pasadena CA 91125; email: vandyk@ipac.caltech.edu.}
\altaffiltext{2}{Department of Astronomy, University of California,
  Berkeley, CA 94720-3411; email: cenko@berkeley.edu,
  afilippenko@berkeley.edu, silverio@astro.berkeley.edu, gmarcy@astro.berkeley.edu, 
  howard@astro.berkeley.edu,
  isaacson@astro.berkeley.edu.}
\altaffiltext{3}{Faculty of Exact Sciences, Tel-Aviv University, 69978 Tel-Aviv-Yafo, Israel;
email: dovi@wise.tau.ac.il.}
\altaffiltext{4}{Department of Particle Physics and Astrophysics, Faculty of Physics, The Weizmann 
Institute of Science, Rehovot 76100, Israel; email: iair.arcavi@weizmann.ac.il, avishay.gal-yam@weizmann.ac.il.}
\altaffiltext{5}{Institute for Astronomy, University of Hawaii, 2680 Woodlawn Dr.,
Honolulu, HI 96822; email: stockton@ifa.hawaii.edu.}
\altaffiltext{6}{Canada-France-Hawaii Telescope Corporation, 65-1238 Mamalahoa Hwy,
Kamuela, HI 96743; email: jcc@cfht.hawaii.edu.}

\begin{abstract}
We report on the 
direct detection and characterization of the probable red supergiant progenitor of 
the intermediate-luminosity Type II-Plateau (II-P) supernova (SN) 2012aw in the nearby (10.0 Mpc) spiral galaxy Messier 95 
(M95; NGC 3351). We have identified the star in both {\sl Hubble Space Telescope\/} images of 
the host galaxy, obtained 17--18 yr prior to the explosion, and near-infrared ground-based 
images, obtained 6--12 yr prior to the SN. The luminous
supergiant showed evidence for substantial circumstellar dust, manifested as excess line-of-sight 
extinction.
The effective total-to-selective ratio of extinction to the star was $R'_V \approx 4.35$, which is 
significantly different from that of diffuse interstellar dust (i.e., $R_V=3.1$), and the total extinction to the 
star was therefore, on average, $A_V \approx 3.1$ mag. 
We find that the observed spectral energy
distribution for the progenitor star is consistent with an effective temperature of 3600~K (spectral type
M3), and that the star therefore had a bolometric magnitude of $-8.29$. 
Through comparison with recent theoretical massive-star evolutionary tracks we can infer that the
red supergiant progenitor had an initial mass $15 \lesssim {\rm M_{ini}} ({\rm M}_{\odot}) < 20$. 
Interpolating by eye between the available tracks, we surmise that the star had initial mass
$\sim 17$--$18\ {\rm M}_{\odot}$. 
The circumstellar dust around the 
progenitor must  have been destroyed in the explosion, as the visual extinction to the SN is
found to be low ($A_V=0.24$ mag with $R_V=3.1$).
\end{abstract}

\section{Introduction}

Supernovae (SNe) are among the most powerful explosions in the Universe. In addition to Type Ia SNe,
which arise from the thermonuclear runaway explosion of a mass-accreting white dwarf star, 
there are SNe that result from the collapse of the core at the endpoint of a massive 
\citep[with initial mass ${\rm M_{ini}} \gtrsim 8\ {\rm M}_{\odot}$; e.g.,][]{woo86} 
star's evolution. If the star explodes with most
of its extended hydrogen envelope still relatively intact, the event will be observed as a Type
II-Plateau SN \citep[SN II-P;][]{barbon79}. We would expect such a progenitor star to be in the red
supergiant (RSG) phase at the time of core collapse.

We have been extremely fortunate in recent years to detect and
characterize the probable RSG progenitors of SNe~II-P in nearby galaxies. 
\citep[We note that the most famous
progenitor identification, of the star Sk $-69\arcdeg$~202 that exploded as SN 1987A in the
Large Magellanic Cloud, was actually a blue supergiant, not a RSG; e.g.,][]{arnett89}.
One of the best examples 
is the identification in high-quality, ground-based imaging data of the RSG progenitor of the 
SN~II-P 2008bk in NGC 7793 \citep{mattila08,vandyk12}. Other SN~II-P RSG progenitors have
also been directly identified in archival {\sl Hubble Space Telescope\/} ({\sl HST}) images of 
nearby host galaxies, including SN 2003gd in M74 \citep{vandyk03,smartt04}, 
SN 2004A in NGC 6207 \citep{hendry06}, SN 2005cs in M51 \citep{maund05,li06},
and SN 2009md in NGC 3389 \citep{fraser11}.
All five of these SNe~II-P are of low luminosity,
with bolometric luminosities $L_{\rm bol} \lesssim 10^{41.5}$ erg s$^{-1}$ at maximum, 
and with lower ejecta velocities
during the plateau and lower luminosity on the light-curve tail as a result of a smaller $^{56}$Ni
yield in the explosion \citep[e.g.,][]{zampieri03,pastorello04}. 
This is relative to intermediate-luminosity 
SNe II-P, such as SN 1999em in NGC 1637 
\citep[e.g.,][]{hamuy01,leonard02,elmhamdi03}, with $L_{\rm bol} \approx 10^{41.5}$--$10^{42}$ 
erg s$^{-1}$ at maximum.

\bibpunct[; ]{(}{)}{;}{a}{}{;}
Three other SNe~II-P also have direct probable progenitor identifications:
SN 1999ev in NGC 4274 \citep{maund05a}, 
SN 2004et in NGC 6946 \citep{li05,crockett11}, 
and SN 2008cn in NGC 4608 \citep{eliasrosa09}. 
SN 1999ev has no published photometry or spectroscopy, so its nature has not been well 
determined.
SN 2008cn appears to have been similar to high-luminosity SNe II-P 
(with $L_{\rm bol} \approx 10^{42.5}$ erg s$^{-1}$ at maximum), such as SN 1992H 
\citep{clocchiatti96} and SN 2007od \citep{inserra11}, 
which also exhibit a less pronounced plateau and more linear post-maximum decline.
Furthermore, 
\citet{eliasrosa09} detected a progenitor for SN 2008cn which was more yellow than red.
\citet{maguire10} showed that SN 2004et bears similarities in
both its photospheric expansion velocity
and the overall shape of its bolometric light curve to those
of SN 1999em (although SN 2004et may have been 
a factor of two more luminous than SN 1999em),
implying that SN 2004et may also have been of intermediate luminosity.
However, the nature and initial mass of the identified progenitor star has been debated
\citep[][Van Dyk \& Jarrett, in preparation]{li05,crockett11}. 
The progenitors of other intermediate-luminosity SNe~II-P, therefore, 
have not yet been directly
identified, although not without efforts in analyzing {\sl HST\/} data to do so in the
recent past, such as for SN 1999em \citep{smartt02}. 

The recent discovery of SN 2012aw in M95 (NGC 3351) by P.~Fagotti on 2012 
March 16.86 (UT dates are used throughout this paper), A.~Dimai on 2012 March 16.84, and J.~Skvarc on March 17.90 (reported in CBET 3054) 
has now afforded us with the best opportunity to do so. 
Indications of the nature of the SN during the plateau
phase, which we will present in a forthcoming paper, are that it is similar to SN 1999em.
A spectrum of the SN on March 17.77 by \citet{munari12} showed a very hot, blue, essentially featureless
continuum. A spectrum on March 18.77 by \citet{siviero12} also showed a blue, featureless
continuum, and later spectra on March 19.85 and 19.92 exhibited the characteristics
of a very young SN~II-P, with the onset of broad lines having P-Cygni-like profiles.
The object was also discovered by the Palomar Transient Factory 
\citep[PTF;][]{law09,rau09} and given the name PTF12bvh.
It was first detected 
at $R \approx 14.2$ mag in an image 
obtained with the PTF camera \citep{rahmer08} on 
the Palomar Oschin Schmidt 48-in (1.2-m) telescope on March 16.70.
\bibpunct[ ]{(}{)}{;}{a}{}{;}
The SN was not visible up to and including March 14.75 to a limiting magnitude 
$R \gtrsim 22$ ($3 \sigma$), providing a tight constraint on the explosion epoch 
\citep[][originally
reported $R \gtrsim 20.7$ mag, but this has since been revised to the fainter limit]{poz12a}. 
\bibpunct[; ]{(}{)}{;}{a}{}{;}
The SN shows indications of interaction with circumstellar matter, through detection in the
radio \citep{yadav12,stockdale12} and X-ray \citep{immler12} bands. It also exhibits signs,
although preliminary, of possibly unusual polarization of the SN ejecta \citep{leonard12}.

The probable progenitor of SN 2012aw/PTF12bvh was first detected in {\sl HST\/} images by
\citet{eliasrosa12} and subsequently by \citet{fraser12a}. The apparently red color of the
star indicated that it was most likely a RSG.
Here we present
photometry of the star and discuss its nature, including its likely initial mass.
An analysis of the progenitor has also been conducted by \citet{fraser12b}.
In their study, \citeauthor{fraser12b} conclude that the RSG was observed through
considerable visual extinction ($A_V > 1.2$ mag), and their estimates of the star's luminosity 
and initial mass span a fairly large range, $10^{5.0}$--$10^{5.6}\ {\rm L}_{\odot}$ and
14--26 M$_{\odot}$, respectively.

For the distance to the SN, we adopt the reddening- and metallicity-corrected distance 
modulus $\mu_0=30.00 \pm 0.09$ mag for M95 determined by \citet{freedman01}.

\section{Progenitor Observations}\label{progobs}

The SN site was imaged by {\sl HST\/} in F439W, F555W, and F814W with the 
Wide-Field Planetary Camera 2 (WFPC2) between 1994 December and 
1995 January, as part of the {\sl HST\/} Key Project to use Cepheid-based distances
to measure the value of the Hubble constant (GO-5397; PI J.~Mould). The Hubble Legacy 
Archive (HLA) project has subsequently combined the individual WFPC2 exposures, using the task
{\it MultiDrizzle} \citep[][see also \citeauthor{fruchter02} \citeyear{fruchter02}]{koekemoer03}, 
into deep image mosaics in F555W and F814W.
The total exposure times are 30130~s and 9830~s in these two bands, 
respectively.
The HLA did not produce mosaics at F439W; the total exposure time is
5000~s for these four images.
The SN site was also imaged in F555W on 1995 December 4 by program GO-5972 
(PI J.~Mould) for 2000~s total, and in F336W and F658N on 2009 January 18 
by program GO-11966 (PI M.~Regan) for 4400~s and 1800~s total, respectively.

Images of the SN were obtained on March 21.31 in the $i'$ band 
in sets of exposures with times of 5~s and 10~s, under very good observing conditions
($\sim 0{\farcs}8$), 
using the MegaCam on the 3.6-m Canada-France-Hawaii Telescope (CFHT),
to initially locate the SN position in the pre-SN
archival {\sl HST\/} images (see below).  
From these images we also measured an absolute position for the SN of
$\alpha$(J2000) = $10^{\rm h}43^{\rm m}53{\fs}73$, 
$\delta$(J2000) = $+11{\arcdeg}40{\arcmin}17{\farcs}9$ ($\pm 0{\farcs}11$ root-mean square),
relative to 12 stars in the field from the USNO B1.0 Catalog \citep{monet03}.
To further refine the SN position, relative to the {\sl HST\/} images, we
obtained three sets of dithered Near-Infrared Camera 2 (NIRC2) images of the SN in the $K_p$ 
band (with exposure times 1.5, 3, and 10~s, respectively), using 
adaptive optics (AO) and the 10-m Keck~II telescope on March 27.42. 
Since the SN was quite bright at the time of these observations, it was possible to perform
the AO using it as a natural guide star, so that the laser guide star was not necessary.

Using the CFHT images, 
we were able to locate a candidate progenitor star for the SN in the HLA F814W image mosaic, 
with an error circle radius of $\sim 0.6$ pixel ($0{\farcs}03$ for these drizzled mosaics), the
results of which were first presented by \citet{eliasrosa12}.
For the Keck AO imaging, from a coaddition of 40
individual 3~s frames (no geometric distortion correction was applied prior to the coaddition)
in which the SN is only mildly saturated, we used 16 stars in common between this coadded
mosaic and the HLA F814W mosaic with the IRAF task {\it geomap} to register the two
mosaics with an uncertainty in the image transformation 
of $\Delta x=0.204$ and $\Delta y=0.591$ HLA mosaic pixels
($0{\farcs}010$ and $0{\farcs}030$, respectively). We find that the location of the SN and
progenitor candidate are consistent to within the $1\sigma$ statistical uncertainty of our
alignment procedure.
\bibpunct[ ]{(}{)}{,}{a}{}{,}
We conclude that we have confirmed the candidate star seen in the WFPC2 mosaics 
as the probable progenitor of SN 2012aw. This is the same star that was identified by \citet{fraser12a, fraser12b}.

Using Dolphot v2.0, as applied to WFPC2 data \citep{dolphin00a,dolphin00b}, we measured
the apparent brightness of the progenitor from the 
ensemble of individual images in both F555W and F814W from GO-5397.
\bibpunct[; ]{(}{)}{;}{a}{}{;}
For F555W we ultimately input into Dolphot only the images which produced uncertainties in 
the photometry of $\lesssim 0.28$ mag; in this band, the star is detected at only 
$\sim 4\sigma$ per individual exposure in the first place. We also omitted one F814W
exposure which was appreciably noisier than all the others (its exposure time was only 230~s,
compared to 1000--1500~s for the other exposures in this band).
We used one of the remaining F814W exposures (the
star is generally 
detected at $\sim 22\sigma$ per exposure) as the astrometric reference image in 
Dolphot. 
The output from Dolphot automatically includes the
transformation from flight-system F555W and F814W to the corresponding
Johnson-Cousins \citep{bessell90} magnitudes in $V$ and $I_C$. (We refer to $I_C$ as $I$
hereafter.)
We find that $m_{\rm F555W} = 26.49 \pm 0.07$ and $m_{\rm F814W}=23.39 \pm 0.02$ mag.
These flight-system magnitudes transform to $V=26.59$ and $I=23.44$ mag.
Although our measurement at F814W is identical to that obtained by \citet{fraser12b}, our
measurement at F555W is brighter by $> 1\sigma$ than their value.
For F336W and F439W the star was not detected by Dolphot to 
$m_{\rm F336W} \gtrsim 22.4$ and $m_{\rm F439W} \gtrsim 25.9$ mag (both $3\sigma$).
The measurements of the progenitor's apparent brightness (or limits to the apparent brightness)
are summarized in Table~\ref{tabprog}.

As \citet{fraser12b} discuss, there are also archival ground-based near-infrared images which
contain the SN progenitor, namely those obtained at the European Southern Observatory (ESO), 
with the Infrared Spectrometer and Array Camera (ISAAC) on the 8.2-m Very Large
Telescope (VLT) Unit Telescope 1 on 2000 March 26 and 27 (PI F.~Bresolin), 
and with the Son of ISAAC (SOFI) on the 3.6-m 
New Technology Telescope (NTT) on 2006 March 24 (PI J.~Ascenso). 
The ISAAC images we 
used were in the $J_s$ band, with 30~s individual frame times and four subintegrations 
(coadditions) in memory, and the SOFI images were in the $K_s$ band, with 8~s 
frame times and 15 subintegrations.
The progenitor site is also in $K_s$ images of the host galaxy, obtained as commissioning
data on 2000 March 23 with the Isaac Newton Group Red Imaging Device (INGRID) on the WHT.
The results of that imaging have been reported by \citet{knapen03}.

\citet{fraser12b} calibrated their photometric analysis of the ISAAC and SOFI 
datasets using 2MASS stars in the images. 
The problem with this approach is that these stars are all near the photometric limit of the 2MASS
survey, $J \approx 16.4$ mag for the ISAAC images and $K_s \approx 14.7$--15.5 mag for 
the SOFI images. At this limit we expect the photometry to be far less reliable, as described in the
2MASS Explanatory 
Supplement\footnote{http://www.ipac.caltech.edu/2mass/releases/allsky/doc/explsup.html.}, and
therefore calibration of the ISAAC and SOFI image data using these stars could suffer from 
systematic effects, leading to a possible error in the final apparent magnitudes for the progenitor.

We therefore reanalyzed the ISAAC and INGRID images, including the original calibration 
data for those observations, and reanalyzed the SOFI images, applying the calibration at $K_s$
from the INGRID observations to those with SOFI, employing five well-detected stars 
in common between the two datasets. This calibration should be valid, since the
bandpasses of both the INGRID and SOFI $K_s$ filters are quite similar.
The data reduction of the raw frames followed standard procedures for near-infrared imaging.
For the ISAAC, SOFI, and INGRID data we first corrected the individual images for the instrumental 
response with
a combined, normalized flat frame obtained in each band. For the ISAAC observations, twilight-sky
frames had been obtained for this purpose; for SOFI observations, dome flats were obtained;
and for the INGRID observations, we produced a flat from
a median-filtered combination of the off-source sky frames.
We then subtracted from the on-source images a median-filtered combination of the sky frames.
For each band a reference frame was chosen, all other frames were shifted in pixel space
relative to the reference, and the shifted frames were coadded, together with the reference frame,
to produce a single image mosaic. 
For the March 26 ISAAC data we coadded 16 individual $J_s$ frames, and for the March 27 we 
coadded 17 frames. For the SOFI $K_s$ imaging we coadded all 5 on-source frames. 
For the INGRID imaging we coadded 9 
of the 20~s frames that contained the progenitor site.

\bibpunct[; ]{(}{)}{;}{a}{}{;}
We extracted photometry from all of the image mosaics using point-spread function (PSF) fitting in 
IRAF/DAOPHOT \citep{ste87}. 
The progenitor star was detected in the $J_s$ mosaics from the two ISAAC nights at
a signal-to-noise ratio ($S/N$) of $\sim 25$--35, 
whereas the star was detected in both the SOFI and INGRID mosaics at only $S/N \approx 4$.
We measured aperture photometry using IRAF for the calibration-star observations obtained for 
both the ISAAC and INGRID runs. For ISAAC these consisted of 
various \citet{persson98} standard stars observed on both nights.
For both 2000 March 26 and 27
we found solutions with airmass corrections of $0.100$--0.105 mag airmass$^{-1}$, 
essentially the
canonical correction at $J$ \citep[][the root-mean square uncertainty in the solution was 0.002 mag for
March 28 and a slightly higher 0.009 mag for March 27]{persson98}.
We therefore measured $J=21.03 \pm 0.04$ mag on March 26 and 
$J=21.01 \pm 0.03$ mag on March 27 for the progenitor.
We note that, although the images are in $J_s$, while the photometric standard magnitudes are in
the Las Campanas Observatory (LCO) $J_{\rm LCO}$ band \citep{persson98}, 
we analyzed the synthetic photometry of three of these standard stars (P041-C, P177-D, and
P330-E), for which there are also {\sl HST\/} calibration spectra, 
and the magnitudes for these stars are only about 0.02
mag brighter through $J_s$ than $J_{\rm LCO}$; we also find that for the likely 
spectral energy distribution (SED) of the probable
progenitor star (see below), the progenitor itself would be about 0.05 mag brighter in $J_s$
than in $J_{\rm LCO}$. 
(We hereafter refer to $J_{\rm LCO}$ merely as $J$.)
We have added this latter difference as an additional uncertainty 
in quadrature with the uncertainty in the photometric measurement.
The uncertainty-weighted mean of the two measurements is $J = 21.02 \pm 0.03$ mag.

For the INGRID calibration, two United Kingdom InfraRed Telescope (UKIRT) faint
standards \citep{leggett06}, FS14 and FS21, had each been observed at the low airmasses 
which bracketed the observations of M95. We obtained aperture photometry of these two stars
and calibrated the faint DAOPHOT photometry of the progenitor star accordingly. 
We note that the $K_s$ bandpass used for these observations 
is in the Mauna Kea Observatory (MKO) system \citep{simons02,tokunaga02}.
The star's brightness in this band is then $K_s = 19.52 \pm 0.27$ mag.
Applying the calibration of the INGRID data to the SOFI image mosaic, as previously described, 
we obtain $K_s = 19.42 \pm 0.29$ mag for the progenitor.
The uncertainty-weighted mean of these two measurements is $K_s = 19.47 \pm 0.19$ mag.

We summarize the measurements of the progenitor's near-infrared brightness in 
Table~\ref{tabprog}. Although, in the end, our measurements at $J$ and $K_s$ agree with those
obtained by \citet{fraser12b} to within the uncertainties in their photometry, we were able to 
achieve a higher precision in our measurements and were able to reduce
the uncertainties substantially through the standard-star calibrations.

\section{The Nature of the Progenitor Star}\label{progenitor}

Unlike the lower-luminosity RSGs which are progenitors of low-luminosity SNe~II-P, such as
SN 2008bk \citep{mattila08,vandyk12}, \citet{massey05} pointed out that 
we would expect higher-luminosity RSGs to lose mass at a 
significantly higher rate, and that this mass loss results in circumstellar 
shells where dust tends to form. 
We already have an indication that this RSG was of high luminosity, from its $K$-band brightness;
even neglecting extinction (which, in the first place, at $K$ is typically nearly a factor of 10 less 
than at $V$) and assuming a bolometric correction $BC_K \approx 3$ mag, 
the bolometric magnitude for the star would be $M_{\rm bol} \approx -7.8$.

What is striking about this progenitor detection is that the star is sufficiently luminous
to be easily visible in both the {\sl HST\/} F814W image and the ground-based near-infrared
images. Another interesting facet is the relative isolation of the progenitor in the host
galaxy; it is apparently far from any noticeable cluster, OB association, or H~{\sc ii} region. 
Most likely, this isolation largely contributes to the detectability of the star at the distance of the 
host galaxy, since the star's environment is uncrowded.
Furthermore, it is tempting to speculate that, like the famous Galactic RSG Betelgeuse 
\citep[$\alpha$ Orionis; e.g.,][]{noriegacrespo97}, the SN 2012aw progenitor could also have been
a stellar runaway.

\subsection{Variability of the Progenitor}

It is known that RSGs can be semiregular or irregular variable stars. 
\citet{kiss06} analyzed the optical 
variability of 48 Galactic RSGs and found evidence for two
modes of variability in 18 of the stars (e.g., $\alpha$ Ori), 
one mode with periods of a few hundred days and one of a few
thousand days. 
Concern therefore should exist whether the apparent brightnesses that we have measured
in the various bands, especially the ensemble from the {\sl HST\/} images, 
are representative of the star's actual brightness.
Whereas variability at $V$ can be considerable, up to a magnitude or more,
RSGs show essentially no variability at $K$ \citep[e.g.,][]{levesque07}.

The {\sl HST\/} data for M95 were obtained over a number of epochs spanning 
$\sim 65$~d (those data were originally taken to discover and 
measure the periods of Cepheids in the host
galaxy), so we can analyze the multi-epoch photometry for any indication of 
short-period variability of the 
progenitor in 1994/1995, $\sim 17$--18 yr prior to the explosion.
In Figure~\ref{figvar} we show the individual measurements with Dolphot from the
{\sl HST}/WFPC2 images in both F555W and F814W. For {\sl HST\/} observations which were
made on essentially the same Julian date, we have computed an uncertainty-weighted mean
value for the star's brightness and show those in the figure, rather than the two individual
measurements for a given date. This was particularly the case for the F814W measurements.
We also show estimates of the color from three epochs.

Clearly excursions of the measurements exist, relative to the mean
brightness for the star in each band, with total 
apparent variations of $\Delta m_{\rm F555W} \approx 1$ mag and
$\Delta m_{\rm F814W} \approx 0.18$ mag. We must determine the significance of these
variations. For that we have computed the reduced $\chi^2$ statistic in each band, and
find for F555W,  $\chi^2_{\rm red}=1.92$, and for F814W, $\chi^2_{\rm red}=5.16$.
We can also assess the $p$ value of these two statistics; a $p$ value of 0.05 or less is usually 
regarded as statistically significant (i.e., the observed deviation from the mean is significant).
For F555W we find $p=0.033$, and for F814W, $p=0.001$. 
We therefore consider it likely that the variability we observe is real. 
We note that \citet{fraser12b} seemingly dismiss the possibility of variability in either band.

The mean color is already $\sim 0.8$ mag redder in $V-I$ than the color of 
an early-M supergiant in the absence of dust. The variation in the star's color is $V-I \approx 0.6$ mag,
relative to the mean. We clearly do not have spectra of the star, so we cannot determine whether
the color variation represents true changes in the star's effective temperature
\citep[e.g.,][]{massey07}, or merely variations in the line-of-sight dust. If the former, the color
variations would correspond to variations in the mean effective temperature of $\sim 300$ K.

\bibpunct[ ]{(}{)}{;}{a}{}{,}
Although the progenitor shows considerable variability over the $\sim 65$~d at $\sim V$, 
which would be consistent with 
variations in the dust content of the star's circumstellar material \citep[e.g.,][]{massey05,massey07}, 
and more limited variability at 
$\sim I$, we will assume that the mean brightness at both $V$ and $I$ over this timescale
is representative of the star's brightness over far longer timescales, at least through the year 2000, 
when the $J$ and $K_s$ brightness were sampled. 
We would expect the $K_s$ brightness to have been relatively 
constant. We have no insight into the level of variability at $J$, but we might expect it to be 
considerably less than at $I$. We therefore assume that the progenitor's observed brightness in
all of these bands fairly represents the actual SED of the star.

\subsection{Metallicity in the SN environment}

Assuming the oxygen abundance can serve as a proxy for the overall metallicity in the host
galaxy, we estimated the metallicity at the SN site based on the gradient in the disk of M95
of the O abundance measured by \citet{moustakas10}.
We deprojected the MegaCam image, assuming the values for the position angle
and inclination for M95 from \citeauthor{moustakas10}, and measured the radial offset of the
SN position from the nuclear position. For a plate scale of $0{\farcs}185$ pixel$^{-1}$,
we find that this offset is $\rho=162{\farcs}02$, or $2{\farcm}70$. 
Again, assuming 
the radial offset at 25 $B$ mag arcsec$^{-2}$ for M95 from \citet{moustakas10}, i.e., 
$\rho_{25}=3{\farcm}71$, we then calculate that $\rho/\rho_{25}=0.73$. 
At this scaled nuclear offset, 
assuming the 
abundance in \citeauthor{moustakas10}~derived from the strong-line index calibration 
by \citet{pt05}, 
we find that 12 + log(O/H) $\approx 8.5$ in the SN environment. 
Given that the solar value is 12 + log(O/H) = $8.66 \pm 0.05$ \citep{asplund05}, 
we infer that the metallicity at the SN site is only slightly subsolar and likely still consistent
with solar, given the uncertainties. We therefore analyze
our results assuming solar metallicity.

\subsection{Properties of the Star}

From the relatively high apparent brightness at $K_s$ for the star, 
compared to the relative faintness at $V$, we realized that the reddening to the progenitor 
had to be high, likely due to circumstellar dust.
\citet{massey05} demonstrated that the effective total-to-selective ratio of absorption, $R'_V$, 
for the dust around Galactic RSGs
should differ considerably from the $R_V=3.1$ typical of the diffuse interstellar medium
\citep[ISM; e.g.,][]{car89}. This may imply a difference, for example, in the grain-size distribution in 
these shells.
They found from their sample of RSGs
that $R'_V = 4.1 + 0.1E(B-V) - 0.2 \log g$, where $\log g$ is the surface gravity of the
RSG and has values of between $-0.5$ and 0.5 (when $g$ is expressed in cgs units). 

We compared the observed SED for the progenitor to a model SED synthesized 
using STSDAS/SYNPHOT within IRAF from the
MARCS model stellar atmospheres for RSGs at solar metallicity \citep{gus08} 
with $T_{\rm eff}$ in the range 3400--3800 K 
\citep[essentially, spectral types of M5 through late K;][]{levesque05},
in steps of 100 K, and surface gravities
$\log g=-0.5$, 0.0, and $+0.5$.
The models are for 15 M$_{\odot}$ stars, assuming spherical geometry
and a microturbulence velocity of 5 km s$^{-1}$. 
The filter response function at $J$ in the LCO system is taken from \citet{persson98}, and the function
for $K_s$ in the MKO system is from the online INGRID instrument 
page\footnote{http://www.ing.iac.es/Astronomy/instruments/ingrid/ingrid{\textunderscore}filters.html.}.
We allowed $R'_V$ to vary from the typical value of 3.1 to larger values and computed a 
range of $A_V$ from 1.4 to 4.2 mag for each $R'_V$, following \citet{car89}.
We were able to eliminate the $\log g=+0.5$ models outright, since none of these were able to 
reproduce the observed SED at any value of $R'_V$, $A_V$, and $T_{\rm eff}$. 
For all of the models at the other two surface gravities, 
we found that only the $T_{\rm eff}=3600$ K models were allowed within the uncertainties
in the photometry. Furthermore,
we found that for the $\log g=-0.5$ models, a range in $R'_V$ from 4.0 to 4.7 was allowed, although
the range in $A_V$ was constrained to 3.05--3.10 mag. For the $\log g=0.0$ models, $R'_V$ 
could range from 4.1 to 4.6, and again, the extinction was constrained to $A_V=3.10$--3.15 mag.

The total $A_V$ in this case can be considered an excess extinction, due to circumstellar dust, 
together with the line-of-sight
extinction to the star, which we assume to be the total extinction to the SN 
(see \S~\ref{sndust}). 
Since we have shown that the star was variable,
possibly due to variations in the amount of circumstellar dust, this total $A_V$ is essentially a
time-averaged value over the duration of both the {\sl HST\/} and ground-based observations.

The average values of $R'_V$ and $A_V$ (in mag) for the models at both surface gravities are
4.35 and 3.10, respectively. 
We show in Figure~\ref{figsed} a representative model at $T_{\rm eff} = 3600$ K with these values 
of $R'_V$ and $A_V$. 
The tightest constraint on the model SEDs 
comes from the $I$-band measurement; the large uncertainty at $K_s$, on
the other hand, is not as constraining.
From \citet{levesque05}, this effective temperature would correspond to spectral type M3.

Following \citet{levesque05} and \cite{bessell98}, we computed the bolometric corrections
at $V$, $I$, and $K_s$ from the MARCS RSG stellar
atmosphere models at $T_{\rm eff}=3600$ K and at $\log g=-0.5$ and $\log g=0.0$.
We found 
$BC_V=-1.79$, $BC_I=0.33$, and $BC_{K_s}=2.79$ mag for $\log g=-0.5$, and
$BC_V=-1.78$, $BC_I=0.32$, and $BC_{K_s}=2.80$ mag for $\log g=0.0$.
Assuming that $R'_V=4.35$ and that $A_V=3.10$ mag, we find that 
$M_{\rm bol}=-8.29 \pm 0.12$, $-8.29 \pm 0.11$, and 
$-8.13 \pm 0.22$ mag from $V$, $I$, and $K_s$, respectively, adding in quadrature the 
uncertainties in the photometric measurements, 
in the inferred extinction (0.05 mag), in the bolometric corrections (0.01 mag), 
and in the host-galaxy distance modulus.
It is very satisfying that our estimates of $M_{\rm bol}$ are exactly the same at $V$ and $I$, which
gives us confidence that the values for $R'_V$, $A_V$, $T_{\rm eff}$, and $\log g$, taken
together, are all consistent. It also allows us to neglect $M_{\rm bol}$ obtained from the far
less certain $K_s$ measurement, although the value of $M_{\rm bol}$ from this band certainly
agrees with those from the other two bands, to within the uncertainties.
We therefore adopt the uncertainty-weighted mean, $M_{\rm bol}=-8.29 \pm 0.08$ mag, 
from $V$ and $I$.
We conservatively adopt an uncertainty of $\pm 100$ K in $T_{\rm eff}$, although
the star's observed SED appears to tightly constrain the effective temperature of the star to less than
this uncertainty. 
Assuming 
$M_{\rm bol} (\sun)$ = 4.74 mag, 
this corresponds to a bolometric luminosity relative to the Sun of 
$\log (L_{\rm bol}/L_{\odot})=5.21 \pm 0.03$. 

In Figure~\ref{fighrd} we show a Hertzsprung-Russell diagram with the locus of the
SN 2012aw progenitor. For comparison we also illustrate the massive-star evolutionary tracks
at solar metallicity from \citet{ekstrom12} for stars with initial rotation which is 40\% of the
critical rotation, at initial masses ${\rm M_{ini}}=15$ and 20 M$_{\odot}$. 
The star is clearly more luminous than the RSG terminus of the ${\rm M_{ini}}=15\ {\rm M}_{\odot}$ 
model, although the star's effective temperature is consistent with the value for the terminus of this 
model. The 20 M$_{\odot}$ model, however, terminates at a higher
luminosity and far hotter $T_{\rm eff}$. The ``red loop'' of that track does approach the star's
locus; however, based on the behavior of that track, we do not  
expect a star with this initial mass prematurely to reach its endpoint at this luminosity and 
cooler effective temperature along the loop.
We can infer, therefore, that the progenitor's
initial mass was in the range $15 \lesssim {\rm M_{ini}} ({\rm M}_{\odot}) < 20$.
\citet{ekstrom12} do not provide model tracks 
between ${\rm M_{ini}}=15$ and 20 M$_{\odot}$, so it is unclear at what luminosity and effective
temperature stars with initial masses within this range would reach their termini. However, by eye
from the figure, we can interpolate that the star's locus could well be consistent with the endpoints
of a putative 17 or $18\ {\rm M}_{\odot}$ track. Clearly, we require the actual model track to be more
precise about this initial mass assignment.

At the adopted effective temperature and luminosity, the star had an effective radius
$R = 1040 \pm 100\ R_{\odot}$. 
From the 
\citet{ekstrom12} ${\rm M_{ini}}=15\ {\rm M}_{\odot}$ evolutionary track, the final
mass is ${\rm M_{fin}}=11.1\ {\rm M}_{\odot}$, so the surface gravity 
would then be $\log g \approx -0.5$. 
From the adopted $A_V$ and $R'_V$, the reddening $E(B-V) = 0.71$ mag.
We then would expect, from the relation from \citet{massey05} between $R'_V$, $E(B-V)$, and 
$\log g$, that $R'_V \approx 4.27$.
This indicates that there is consistency among these three parameters taken together, and 
this cross-check further provides us with confidence in our estimates of the star's properties.

We note that the main difference between the analysis we have presented here and that 
presented by \citet{fraser12b} --- namely, that we are able to better constrain the effective temperature,
luminosity, and, therefore, the initial mass estimate for the progenitor ---
stems not only from the differences in the photometry between the two studies (we found that the
star is brighter in $V$, and we were also able to reduce the uncertainties at $J$ and $K_s$), but also
from our assumption that $R_V$ for the RSG progenitor was different from the typical value of 3.1 for
the diffuse ISM. 
One item to also note is that the evolutionary tracks \citep[without rotation;][]{eldridge08} 
employed by \citet{fraser12b} tend to terminate at significantly cooler effective 
temperatures than the \citet{ekstrom12} tracks that we have used.
The key, ultimately, was connecting the inferred properties of the SN 2012aw progenitor, based on 
our measurements, to those of the Galactic RSGs of similar luminosity and mass \citep{massey05}.

\subsection{The Dust Around the Progenitor}

As \citet{massey05} discussed, 
we can estimate the mass of the dust and duration of dust 
production that is responsible for the excess extinction which we infer for the RSG 
progenitor. \citeauthor{massey05}~assume a thin-shell approximation for the excess extinction, 
$\Delta A_V = \Delta{R} (3.2 \times 10^3) {\rm M}_d/(4\pi R^2 \Delta{R})$, where $R$ is the stellar
radius (in m), $\Delta{R}$ is the extent of the thin dust layer or shell (or the path length through
the dust) above the stellar surface, and ${\rm M}_d$ is the dust mass (in kg). 
The circumstellar matter presumably extended several stellar radii above the star's surface, 
as in the case of $\alpha$ Ori, 
and the overall mass loss could well have been driven by convection in the envelope
\citep{josselin07,chiavassa10}.
It is in the last $\sim 1400$~yr of the RSG phase, as seen in the $15\ {\rm M}_{\odot}$ model from
\citet{ekstrom12}, 
when the star's luminosity, radius, and total mass loss particularly increase.
From Figure 4 of \citet[][]{massey05}, we see that the inferred $\Delta A_V \approx 3$ mag 
(we will show in \S~\ref{sndust} that the interstellar extinction, for $R_V=3.1$, is likely 
only $A_V \approx 0.2$ mag)
is entirely consistent with the $M_{\rm bol}$ of the star. The dust production rate, 
$\dot {\rm M}_d$,
corresponding to this luminosity is $\sim 10^{-8.44}\ {\rm M}_{\odot}$ yr$^{-1}$. 
From the relation above, the value of ${\rm M}_d$ is $\sim 6.1 \times 10^{21}$ kg.
Comparing this to $\dot {\rm M}_d \Delta t$, we find that a dust-producing episode of 
$\Delta t \approx 1$ yr, at some point prior to 1994 
(since this dust already existed by the time of the
first {\sl HST}/WFPC2 images), could account for $\Delta A_V$. 
Although quite a short interval of time,
it is consistent with that inferred for episodic dust production in Galactic RSGs 
\citep{danchi94,bester96}. 

\subsection{The Dust Around the Supernova}\label{sndust}

However brief was the dust production, as \citet{fraser12b} point out, the dust was far more 
quickly destroyed, likely by the X-ray/UV flash within hours of core collapse. As also noted by
\citet{fraser12b}, 
such circumstellar dust destruction is not unprecedented for other SNe 
\citep[e.g.,][]{dwek08,wesson10}.

\citet{poz11} have found that the Na~{\sc i}~D feature strength in low-resolution optical SN
spectra is a poor indicator of the amount of extinction to the SN. 
However, \citet{poz12b} 
have established a well-calibrated relation between reddening and
the equivalent width (EW) of the Na~{\sc i}~D doublet, D1 and D2, 
based on more than one hundred high-resolution spectra of objects through a number
of interstellar lines-of-sight. We
have therefore measured the EW of the doublet, from both the Milky Way and host-galaxy 
components, as clearly detected in a high-resolution
spectrum of SN 2012aw we obtained with the High Resolution Echelle Spectrometer 
\citep{vogt94} 
on the 10 m Keck~I telescope in the red optics configuration (``HIRESr'') on 2012 April 10.29. 
We used the C2 decker (i.e., the $0{\farcs}86$ slit), providing coverage of $\sim$
3800--7300 \AA\ with a resolution of 50,000. 
The portion of this spectrum, centered around the Na~{\sc i}~D feature,  
is shown in Figure~\ref{fighires}. 
We find that for the Milky Way component, EW(D2\,$\lambda$5891.41) =
$94 \pm 8$ m\AA\ and EW(D1\,$\lambda$5897.39) = $56 \pm 9$ m\AA.
For the component internal to M95, EW(D2\,$\lambda$5909.33) = $269 \pm 14$ m\AA\ and
EW(D1\,$\lambda$5915.32) = $231 \pm 11$ m\AA. Uncertainties in the centroids of the 
absorption features are typically $\sim 0.05$ \AA. (These are all vacuum wavelengths, 
corrected to a heliocentric frame of reference.) The features due to M95, particularly, the D2
component and, to a lesser extent, the D1 component, may be slightly saturated.

From the \citet{poz12b} relations, $E(B-V)=0.022 \pm 0.013$ mag from the Milky Way and
$E(B-V)=0.055 \pm 0.014$ mag for the host galaxy. 
(The uncertainties here are primarily from the systematic uncertainties in the relations; the
measurement uncertainties are comparatively negligible.) These relations take into account
saturation in the features, which is a smaller effect compared to the systematic uncertainties in 
the relations.
Our reddening estimate is consistent with the Galactic foreground reddening estimate from 
\citet{sch98}, $E(B-V)=0.028$ mag.
The total reddening indicated from the high-resolution SN spectrum, assuming $R_V=3.1$, 
is then $E(B-V)=0.077$ mag, which is relatively low and comparable to 
the estimated $E(B-V)=0.1$ mag found for SN 1999em \citep{baron00,leonard02,elmhamdi03}.
The visual extinction to the SN is then $A_V=0.24$ mag.
Although \citet{poz09} found that for a sample of SNe~II-P the best-fit $R'_V \approx 1.7$, the 
relations from \citeauthor{poz12b} likely do not depend strongly on $R_V$, and therefore the 
reddening and extinction values we have estimated likely will not be significantly different if
$R_V \neq 3.1$, especially local to the SN. 
Nonetheless, it is evident that the SN explosion must have
destroyed much or all of the circumstellar dust around the progenitor, leaving only what is
most likely interstellar line-of-sight extinction, which we measured from the high-resolution SN 
spectrum.

\section{Conclusions}\label{conclusions}

The probable progenitor of the intermediate-luminosity SN~II-P 2012aw/PTF12bvh has been 
identified in archival {\sl HST\/} optical and ground-based near-infrared images. Using the photometry
extracted from those images we have constructed a SED for the star, and we analyze
the SED to show that the star was a luminous ($M_{\rm bol} = -8.29$ mag) RSG with 
spectral type M3 ($T_{\rm eff}=3600$ K) and with substantial circumstellar dust 
up to 18~yr before explosion. This dust was clearly destroyed by the explosion, since the current
extinction to the SN is relatively low. Although the existing, state-of-the-art, theoretical 
stellar evolutionary tracks do not terminate at the locus of the star in the Hertzsprung-Russell diagram, 
we surmise that the star had an initial mass $\sim 17$--$18\ {\rm M}_{\odot}$.

This mass is essentially the same as the upper limit to the initial mass, 
$16.5 \pm 1.5\ {\rm M}_{\odot}$, that
\citet{smartt09} derived for SN~II-P progenitors.
It is not evident whether this result for SN 2012aw
can be generalized for other intermediate-luminosity SNe~II-P.
The initial mass for the progenitor of SN 2004et is still a subject of debate
\citep[e.g.,][]{li05,crockett11}.
Additionally, 
the ($7\sigma$) upper limit to the detection of SN 1999em at $I=22.0$ mag from 
\citet{smartt02}, at a distance of 11.7 Mpc 
(\citealt{leonard03}, which implies a luminosity limit only a 
factor $\sim 1.4$ fainter than if SN 1999em were at 10.0 Mpc, as is SN 2012aw), 
precludes detection of an analog to the SN 2012aw progenitor. 
It therefore remains uncertain what is the upper limit on the initial mass of
the RSGs that give rise to ``normal,'' intermediate-luminosity SNe~II-P.
The detected, unusual progenitors of high-luminosity SNe~II-P 
\citep[e.g., SN 2008cn;][]{eliasrosa09} may provide some indication of this limit. 
The recent stellar evolutionary tracks from \citet{ekstrom12}, as well as those with
pulsationally driven superwinds by \citet{yoon10}, demonstrate the role of rotation and
mass loss in the late-stage evolution of stars with $M_{\rm ini} \gtrsim 20\ {\rm M}_{\odot}$. 
Still to be investigated more fully are the influences of factors such as
the metallicity and
binarity \citep[e.g.,][]{smith11}.
To verify the candidate SN 2012aw progenitor, we will need to return at very late times, when the 
SN has substantially faded, to see whether the dusty RSG has vanished.

\acknowledgements

We thank the referee for useful comments that helped improve this manuscript.
This work is based in part on observations with the NASA/ESA {\sl
  Hubble Space Telescope}, obtained at the Space Telescope Science
Institute (STScI), which is operated by AURA, Inc., under NASA
contract NAS5-26555.  We thank Nancy Elias-Rosa for initial analysis
of the {\sl HST\/} and CFHT images, and Mark Sullivan for information
about the upper limit to the SN 2012aw/PTF12bvh detection by PTF. Some
of the data presented herein were obtained at the W. M. Keck
Observatory, which is operated as a scientific partnership among the
California Institute of Technology, the University of California, and
NASA; the observatory was made possible by the generous financial
support of the W. M. Keck Foundation.  A.G. is supported by grants
from the ISF, BSF, and GIF foundations, a Weizmann Minerva grant, and
the Lord Sieff of Brimpton Fund.  The research of A.V.F. and his group
at UC Berkeley is funded by Gary and Cynthia Bengier, the Richard and
Rhoda Goldman Fund, NSF grant AST-0908886, the TABASGO Foundation, and
NASA/HST grant AR-12623 from STScI.  This work was supported in part
by the NSF grant PHY-1066293 and the hospitality of the Aspen Center
for Physics.  We dedicate this paper to the memory of our dear friend
and colleague, Weidong Li, with whom two of us (A.V.F. and S.D.V.)
spent many fun years identifying the progenitor stars of SNe; his
tragic passing deeply saddened those who knew him.

\begin{deluxetable}{cc}
\tablewidth{1.5truein}
\tablecolumns{2}
\tablecaption{Photometry of the SN 2012aw Progenitor\tablenotemark{a}\label{tabprog}}
\tablehead{
\colhead{Band} & \colhead{Magnitude} }
\startdata
F336W & $\gtrsim 22.4$\tablenotemark{b} \\
F439W & $\gtrsim 25.9$\tablenotemark{b} \\
F555W & 26.49(07) \\
$V$ & 26.59(07) \\ 
F814W  & 23.39(02) \\
$I$ & 23.44(02) \\
$J$ & 21.02(03) \\
$K_s$ & 19.47(19) \\
\enddata
\tablenotetext{a}{Uncertainties $(1\sigma)$ 
are given in parentheses, in units of 0.01 mag.}
\tablenotetext{b}{These are $3\ \sigma$ upper limits.}
\end{deluxetable}

\clearpage

\begin{figure}
\figurenum{1}
\includegraphics[angle=0,scale=0.70]{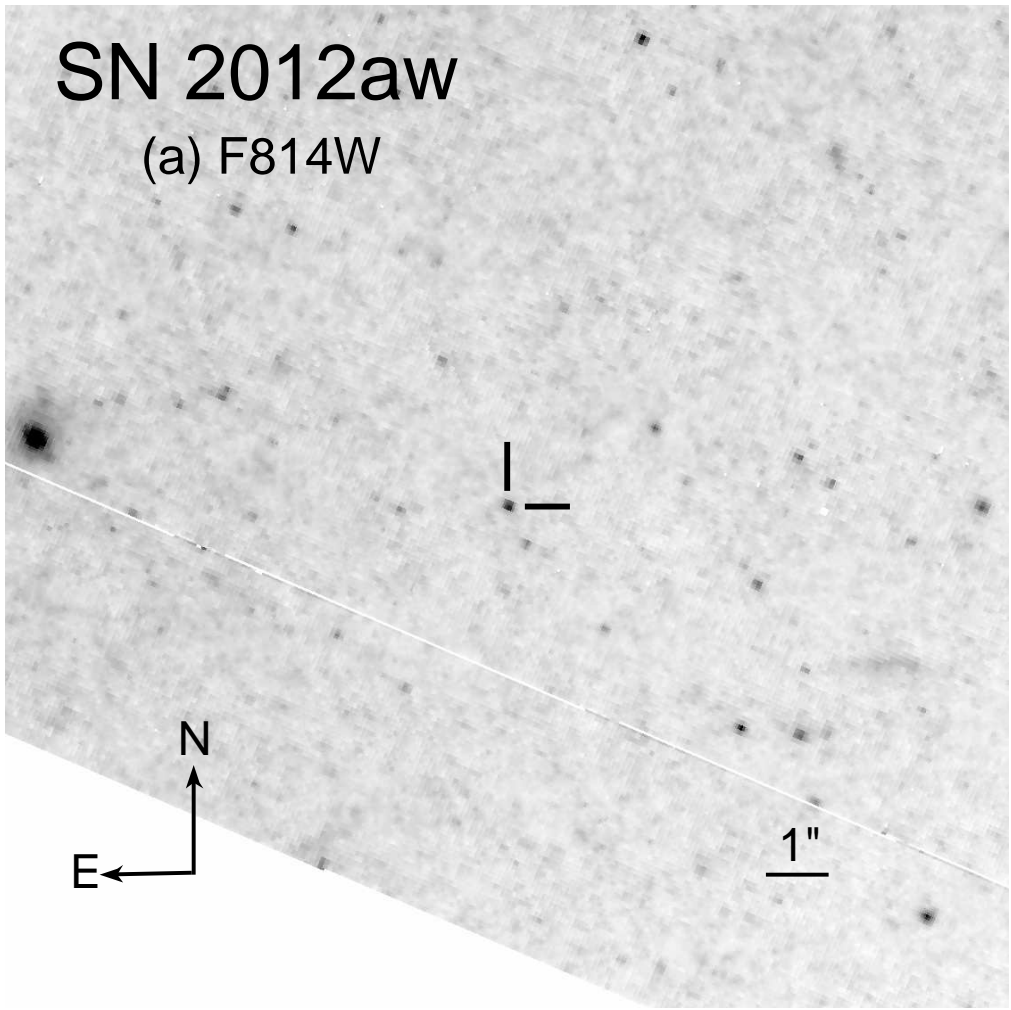}
\caption{(a) A portion of the archival {\sl HST\/} F814W image of M95 from 1994/1995; the
star detected at the precise location of SN 2012aw is indicated by {\it tick marks}. The SN site is
very near to the edge of the image mosaic.
(b) A portion of the $K_p$-band AO image
obtained using NIRC2 on the Keck-II telescope on 2012 March 27.\label{figprog}}
\end{figure}

\clearpage

\begin{figure}
\figurenum{1}
\includegraphics[angle=0,scale=0.70]{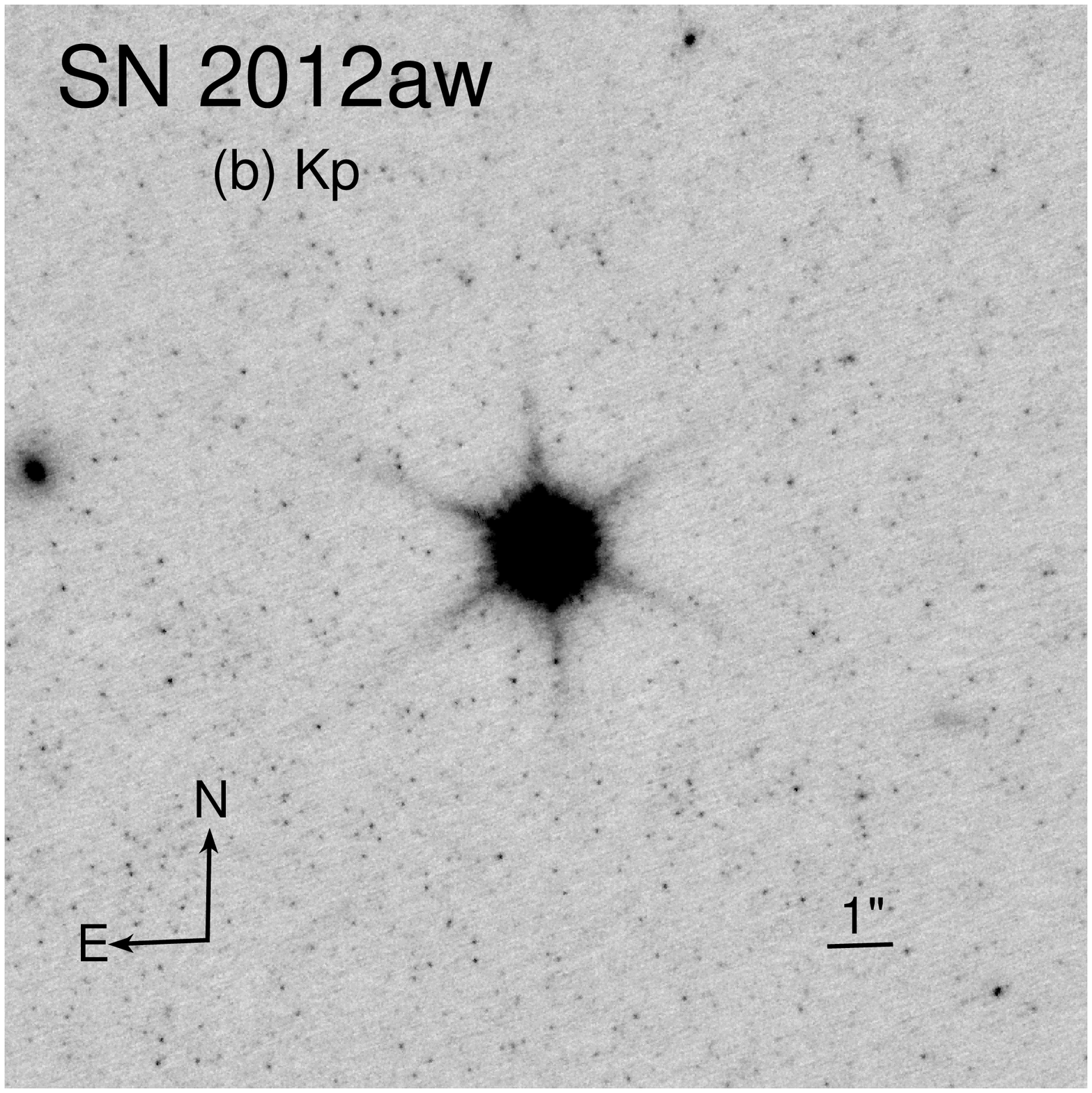}
\caption{(Continued.)}
\end{figure}

\clearpage

\begin{figure}
\figurenum{2}
\includegraphics[angle=0,scale=0.70]{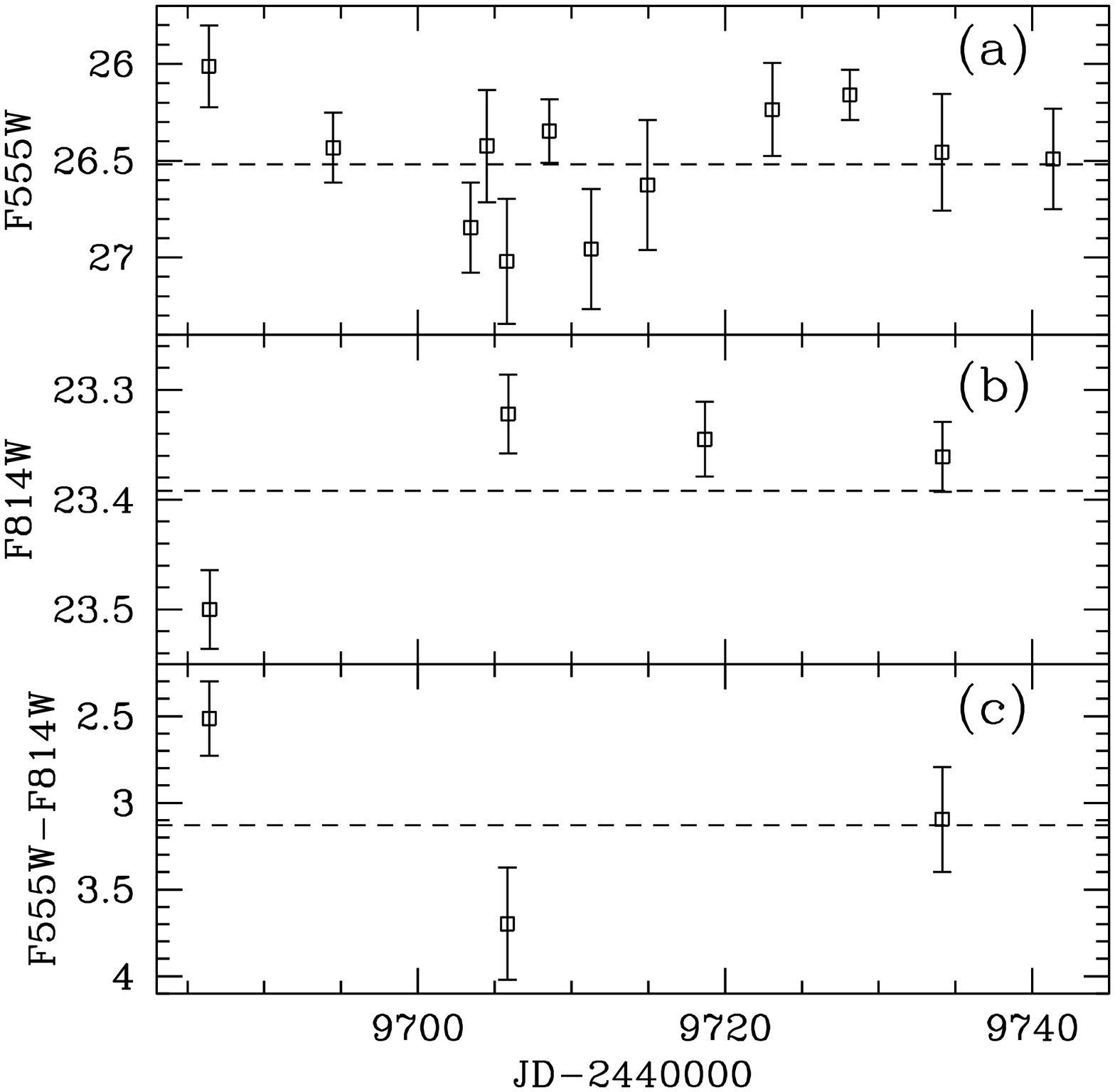}
\caption{Photometry (mag) of the SN 2012aw progenitor measured with Dolphot from the individual
{\sl HST}/WFPC2 observations from program GO-5397 in the (a) F555W and (b) F814W
bands. Observations obtained on the same date in a given band have been averaged, weighted by
the uncertainties in the individual measurements. Panel (c) shows the F555W$-$F814W color (mag).
The {\it dashed lines\/} in all three panels arise from 
the uncertainty-weighted means of all measurements in each band, as returned by
Dolphot.\label{figvar}}
\end{figure}

\clearpage

\begin{figure}
\figurenum{3}
\includegraphics[angle=0,scale=0.70]{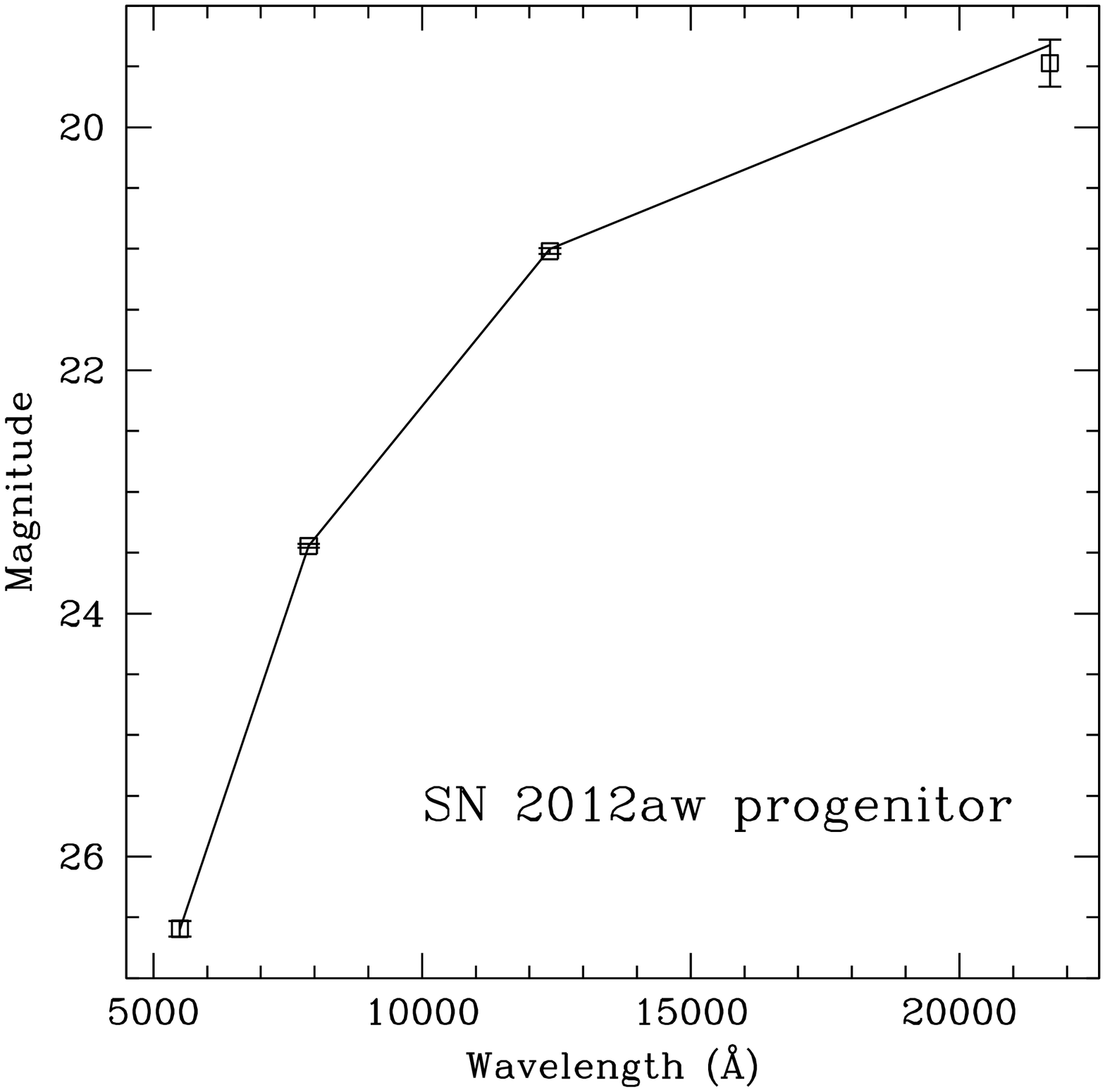}
\caption{The observed SED for the probable SN 2012aw progenitor. See Table~\ref{tabprog}.
Also shown for comparison is an example SED computed via synthetic photometry of a MARCS 
model RSG stellar atmosphere \citep{gus08} with surface gravity $\log g=-0.5$, 
effective temperature $T_{\rm eff}=3600$ K, and extinction $A_V=3.10$ mag
({\it solid line}), assuming an
effective total-to-selective ratio of absorption, $R'_V=4.35$. 
Although a surface gravity $\log g=0.0$ and a limited range in $R'_V$ are also allowed, both 
$A_V$ and
$T_{\rm eff}$ are constrained by the observations; see the text. The model SED
has been normalized at $I$, the band which provides the tightest photometric constraint.
\label{figsed}}
\end{figure}

\clearpage

\begin{figure}
\figurenum{4}
\includegraphics[angle=0,scale=0.70]{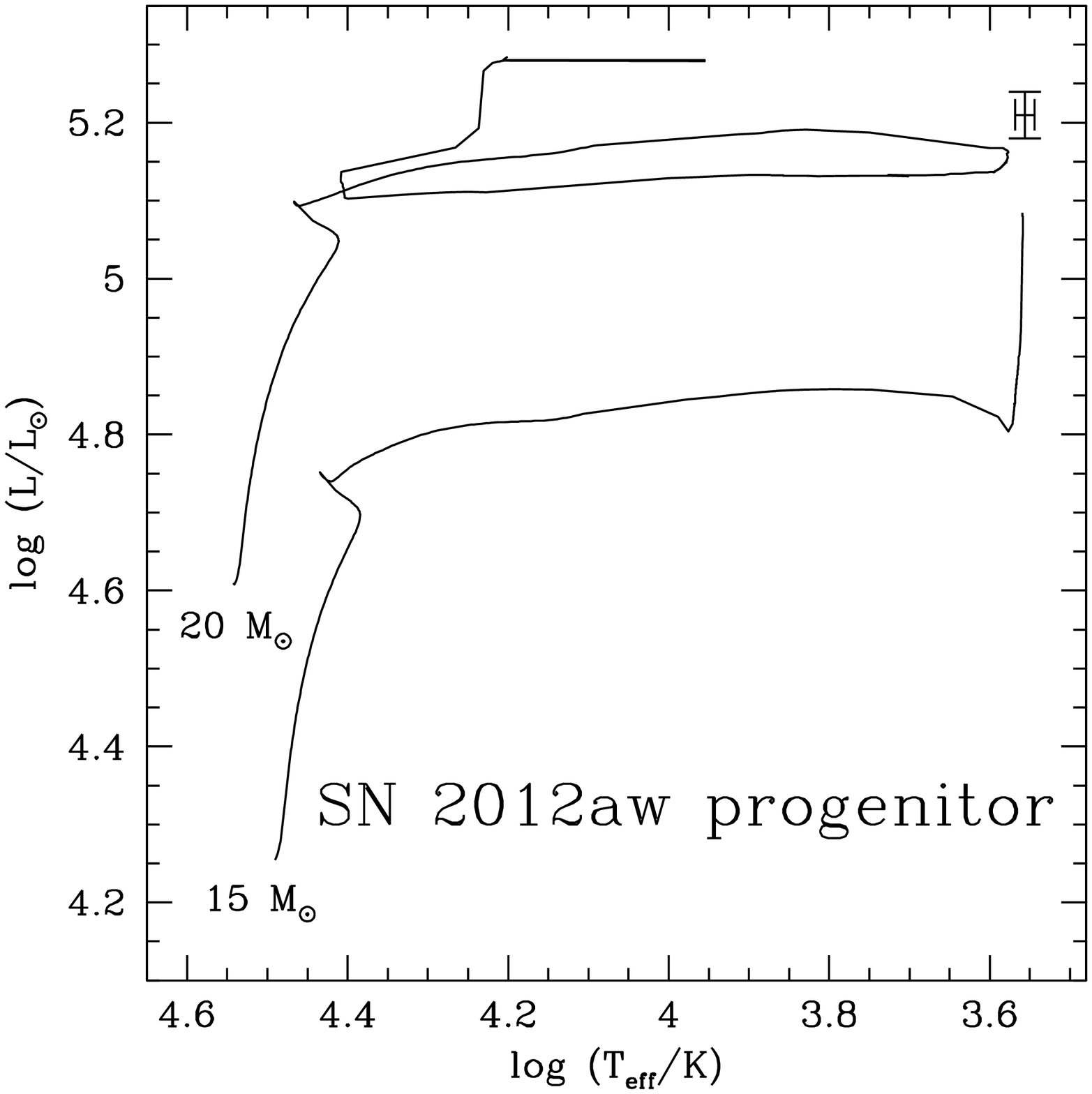}
\caption{Hertzsprung-Russell diagram showing the locus of the SN 2012aw progenitor.
For comparison we also indicate massive-star evolutionary tracks at solar metallicity, which
include initial rotation on the main sequence that is 0.4 times the critical rotation velocity,
from \citet{ekstrom12} for ${\rm M_{ini}} = 15$ and 20 M$_{\odot}$.\label{fighrd}}
\end{figure}

\clearpage

\begin{figure}
\figurenum{5}
\includegraphics[angle=0,scale=0.45]{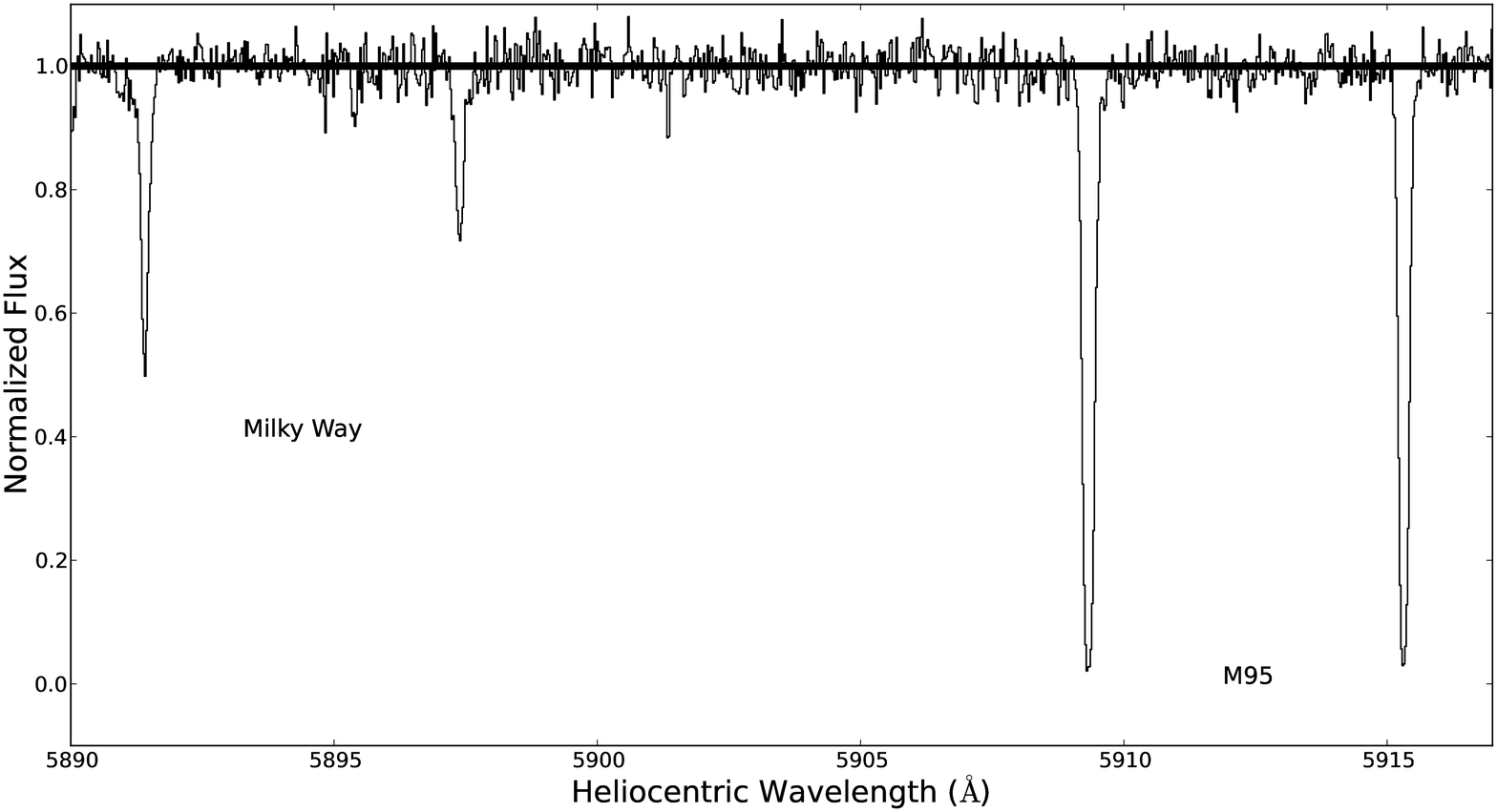}
\caption{Spectrum of SN 2012aw obtained on 2012 April 10.29 with
HIRESr on the 10~m Keck I telescope, centered on the Na~{\sc i} D absorption feature. 
Both feature components D1\,$\lambda$5896 \AA\ and D2\,$\lambda$5890 \AA\ are clearly 
detected from both the Galactic foreground
(labeled as ``Milky Way'') and the SN host galaxy (labeled as ``M95''). 
The features due to the host may be partially saturated.\label{fighires}}
\end{figure}

\end{document}